\documentclass[aps,prl,reprint,superscriptaddress,showpacs,nofootinbib]{revtex4-1}

\usepackage{ulem}
\usepackage{graphicx}
\usepackage{color}
\usepackage{amsfonts}

\usepackage{hyperref}

\renewcommand{\d}{{\rm d}}

\newcommand{\lamu}{\lambda_{\text{u}}}
\newcommand{\Nu}{N_{\text{u}}}
\newcommand {\Omu}  {{\Omega_{\rm u}}}

\newcommand {\E}  {\varepsilon}
\newcommand {\om} {\omega}
\newcommand {\Om} {\Omega}

\begin{document}

\title{Extremely brilliant crystal-based light sources}

\author{Gennady B. Sushko}
\affiliation{MBN Research Center, Altenh\"{o}ferallee 3, 60438 Frankfurt am Main, Germany}

\author{Andrei V. Korol}
\email[]{korol@mbnexplorer.com}
\affiliation{MBN Research Center, Altenh\"{o}ferallee 3, 60438 Frankfurt am Main, Germany}

\author{Andrey V. Solov'yov}
\email[]{solovyov@mbnresearch.com}
\affiliation{MBN Research Center, Altenh\"{o}ferallee 3, 60438 Frankfurt am Main, Germany}

\begin{abstract}
Brilliance of novel gamma-ray Crystal-based Light Sources (CLS) 
that can be constructed through exposure of oriented crystals  
to beams of ultrarelativistic charged particles
is calculated basing on the atomistic scale numerical modeling of the channeling process.
In an exemplary case study, the brilliance of radiation emitted in a diamond-based Crystalline Undulator 
LS by a 10 GeV positron beam available at present at the SLAC facility is computed.
Intesity of CU radiation in the photon energy range $10^0-10^1$ MeV, which is inaccessible to 
conventional synchrotrons, undulators and XFELs, greatly exceeds that of 
laser-Compton scattering LSs and can be higher than predicted in the Gamma Factory proposal to CERN.
Construction of novel CLSs is a challenging task which constitutes a highly interdisciplinary field 
entangling a broad range of correlated activities.
CLSs provide a low-cost altenative to conventional LSs and have enomorous number of applications.

\end{abstract}


\maketitle

 
Development of light sources (LS) operational at wavelengths $\lambda$ well below one angstrom 
(corresponding photon energies $\hbar\om > 10$ keV) is a challenging goal of modern physics. 
Sub-angstrom wavelength, ultrahigh brilliance, tunable LSs will have a broad range of exciting potential 
cutting-edge applications. 
These applications include exploring elementary particles, probing nuclear structures and
photonuclear physics, and examining quantum processes, which rely heavily on gamma-ray sources in
the MeV to GeV range
\cite{KorolSolovyov:EPJD_CLS_2020,Zhu-EtAl:ScieAdv_vol6_eaaz7240_2020,NextGenerationGammaRayLS2020}. %
Modern X-ray Free-Electron-Laser (XFEL) can generate X-rays with wavelengths $\lambda\sim1$ \AA{}  
\cite{Seddon-EtAl_RepProgPhys_v80_115901_720_2017,Doerr-EtAl_NatureMethods_v15_p33_2018,%
SwissFEL_ApplScie_v7_720_2017, Bostedt-EtAl_RMP_v88_015007_2016}.
Existing synchrotron facilities provide radiation of shorter wavelengths but orders of magnitude
less intensive
\cite{Couprie:JElSpectRelPhen_v196_p3_2014,Tavares-EtAl:JSynchrRad_v21_p862_2014,%
YabashiTanaka_NaturePhotonics_v11_p12_2017}.
Therefore, to create a powerful LS in the range $\lambda \ll 1$ \AA{} new approaches and technologies
are needed.

\textcolor{black}{
In recent years \cite{KorolSolovyov:EPJD_CLS_2020,KorolSolovyov:EPJD_Colloquium_2021} significant
efforts of the research and technological communities have been devoted to design and practical
realization of novel gamma-ray Crystal-based LSs (CLS) that can be set up by exposing oriented
crystals to beams of ultrarelativistic positrons or electrons.
Manufacturing of CLSs is a subject of the current European Horizon-2020 project
'N-LIGHT'  \cite{N-Light} and upcoming Horizon-EIC-2021 project TECHNO-CLS.
Construction of a CLS is a challenging technological task, which requires a highly interdisciplinary approach combining development of technologies for crystalline
sample preparation (engaging material science, nanotechnology, acoustics, solid state physics) together with a detailed experimental programme (for CLS characterization, design of incident particle beams, experimental characterization of the radiation emitted) as well as theoretical analysis and advanced  computational modelling for the visualisation and characterisation of CLSs
based on an atomistic level description.
}

\textcolor{black}{
This Letter provides accurate predictions for the brilliance of a Crystalline Undulator (CU) \cite{ChannelingBook2014}
LS on the basis of all-atom molecular dynamics simulations of relativistic particles channeling and radiation in oriented crystals.
The exemplary case study presented below shows that using currently available positron beam it is realistic to achieve the brilliance of radiation emitted in the
photon energy range $\hbar \om \gtrsim 1$ MeV
that exceed brilliance of the laser-Compton scattering LSs
\cite{Rehman_EtAl_ANE_v105_p150_2017,Kraemer_EtAl-ScieRep_v8_p139_2018,
Krafft-Priebe:RevAccScieTechnol_vol3_p147_2010,%
Sei-EtAl:ApplSci_vol10_p1418_2020,NextGenerationGammaRayLS2020,Wu-EtAl:PRL_vol96_224801_2006}.
and of the LS  well as of the predicted in the Gamma Factory proposal for
CERN \cite{Krasny:2018xxv,GammaFactory:Letter_of_intent2019}.
}

\textcolor{black}{
Numerical modeling of the channeling and related phenomena beyond the continuous potential framework
has been carried out by means of the multi-purpose computer package \textsc{MBN Explorer}
\cite{MBN_Explorer_2012,MBN_ChannelingPaper_2013,MBNExplorer_Book} and  a supplementary special multitask
software toolkit \textsc{MBN Studio} \cite{MBN_Studio_2019}.
A special module of \textsc{MBN Explorer} allows one to simulate the motion of relativistic projectiles along with
dynamical simulations of the environment \cite{MBN_ChannelingPaper_2013}.
The uniqueness of the computation algorithm is that it accounts for the interaction of projectiles with all atoms of the
environment thus making it free of simplifying model assumptions.
In addition, a variety of interatomic potentials implemented facilitates rigorous simulations of various media,
a crystalline one in particular.
Overview of the results on channeling and radiation of charged particles in oriented linear, bent and periodically bent crystals
simulated by means of  \textsc{MBN Explorer} and  \textsc{MBN Studio} can be found in
\cite{KorolSolovyov:EPJD_CLS_2020,KorolSolovyov:EPJD_Colloquium_2021,ChannelingBook2014,MBNExplorer_Book}.
}

\textcolor{black}{In Ref. \cite{KorolSolovyov:EPJD_CLS_2020} brilliance of radiation emitted in the photon energy range  $10^0$-$10^1$  MeV in a CU-LS has been estimated
using the model based the continuous interplanar potential concept \cite{Lindhard}
and utilizing the phenomenological approach to describe the dechanneling process.
The model has allowed one to establish optimal parameters of a periodically bent crystal (these include crystal thickness $L$ in the
direction of beam, - the $z$-direction,
bending amplitude $a$ and period $\lamu$) that ensure the highest values of brilliance of the CU-LS for a positron beam of given energy $\E$, transverse beam sizes $\sigma_{x,y}$ and angular divergence $\sigma_{\phi_{x,y}}$.}

\textcolor{black}{The results presented in this Letter have been obtained for a $\E=10$ GeV
positron beam propagating
through an oriented periodically bent diamond (110) crystal.
The beam sizes and divergence used in the simulations are
$\sigma_{x,y} = 32,\, 10$ microns and $\sigma_{\phi_{x,y}}=10,\, 30$ $\mu$rad, respectively.
They correspond to normalized emittance
$\gamma\epsilon_{x,y}=\gamma\sigma_{\phi_{x,y}}\sigma_{x,y}=6.3,\, 5.9$ m-$\mu$rad
and are within the ranges indicated for the FACET-II beam available at the SLAC facility \cite{FACETII_Conceptual_Design_Rep-2015}
(see Supplemental Material (SM)).}

\begin{table}[h]
\caption{
Two sets of parameters of diamond-based CUs used in the simulations:
bending amplitude $a$,  period $\lamu$,
number of periods $N_{\rm u}$,
crystal thickness $L=N_{\rm u}\lamu$.
The last column presents the energies $\hbar\om_0$ of
the fundamental harmonic in the
forward direction.
}the relativistic Lorentz factor
\centering
\begin{tabular*}{8.2cm}{@{}rrrrrr}
\hline
Set  \hspace*{0.25cm}& $a$ (\AA) \hspace*{0.25cm}  &$\lamu$ ($\mu$m) \hspace*{0.25cm}
& $N_{\rm u}$ \hspace*{0.25cm} & $L$ (mm) \hspace*{0.25cm}  & $\hbar\om_1$ (MeV) \\
\hline
(I)  \hspace*{0.25cm}& 20.9 \hspace*{0.25cm}  &   85   \hspace*{0.25cm} &  85         \hspace*{0.25cm} & 7.06 \hspace*{0.25cm} &  2.0  \\
(II) \hspace*{0.25cm}&  5.3 \hspace*{0.25cm}  &   38   \hspace*{0.25cm} & 180         \hspace*{0.25cm} & 6.84 \hspace*{0.25cm} & 10.0  \\
\hline
\end{tabular*}
\label{Table01}
\end{table}

\textcolor{black}{
The values of bending amplitude $a$, and period $\lamu$ as well as the number of periods
$N_{\rm u}$ and the crystal thickness $L$ used in the simulations are listed in Table \ref{Table01}.
Within the framework of the model approach \cite{KorolSolovyov:EPJD_CLS_2020} it has been established that each set of these parameters provides the maximum brilliance of the CU-LS radiation emitted in the forward
direction at the fundamental harmonic energy $\hbar\om_0$ indicated in the last column.
The frequency $\om_0$ is calculated as follows
\begin{eqnarray}
\om_{0} = {2\gamma^2 \Omu \over 1 +  K^2/2}
\label{eq.01}
\end{eqnarray}
where $\gamma$ is the relativistic Lorentz factor,
$K=2\pi\gamma a/\lamu$ stands for an undulator parameter
and $\Omu=2\pi c/\lamu$.
}

\textcolor{black}{
It is important to mention that the values of $a$, $\lamu$ and $L$ indicated in the table
are accessible by means of existing modern technologies.
Construction of novel gamma-ray CLSs is a challenging task involving a broad range of correlated research and technological activities.
One of the key technological task concerns manufacturing of crystals of different desired geometry.
High quality of the undulator material is essential for achieving strong effects in the emission spectra.
A systematic review of different technologies exploited for manufacturing of crystals of different
type, geometry, size, quality, etc. one finds in Refs.
\cite{KorolSolovyov:EPJD_CLS_2020,KorolSolovyov:EPJD_Colloquium_2021}.
A brief summary is presented in SM.
}

To simulate the trajectories of the beam particles in the crystalline medium, the $y$-axis
was chosen along the $\langle 110\rangle$ axial direction. 
This choice ensures that a sufficiently big fraction $\xi$ of the incident beam 
is accepted in the channeling mode at the crystal entrance. 
For a Gaussian beam this fraction can be estimated as 
$\xi=(2\pi\sigma_{\phi_y}^2)^{-1/2}\int_{-\Theta_{\rm L}}^{\Theta_{\rm L}}\exp\left(-\phi^2/2\sigma_{\phi_y}^2\right)\d\phi$
where $\Theta_{\rm L}$ stands for Lindhard's critical angle.
Using $U_0\approx20$ eV for the interplanar potential depth in diamond(110) one calculates $\Theta_{\rm L} =(2U_0/\E)^{1/2} \approx 63$ $\mu$rad
that exceeds by a factor of two the divergence $\phi_y=30$ $\mu$rad. 
The beam divergence $\phi_x=10$ $\mu$rad along the $x$ transverse direction is much smaller than
the natural emission angle $\theta_\gamma=\gamma^{-1} \approx 50$ $\mu$rad. 
As a result, one can expect that big fraction of radiation emitted by the channeling particles 
will be collected within the cone $\theta_0\leq \theta_\gamma$ centered along the incident beam.

The simulations performed aimed at providing accurate quantitative data 
on the brilliance of a CU-LS.
Brilliance, $B$, of a LS is proportional to the number of photons $\Delta N_{\om}$ of 
frequency within the interval $[\om-\Delta\om/2,\om+\Delta\om/2]$ emitted in 
the cone $\Delta\Omega$ per unit time interval, 
unit source area, unit solid angle and per a bandwidth (BW) $\Delta\om/\om$ 
\cite{SchmueserBook}. 
To calculate this quantity it is necessary to know the beam electric current $I$,
its transverse sizes and divergences as well as
the divergence angle $\phi=\sqrt{\Delta\Omega/2\pi}$ and the 'size' 
$\sigma=\lambda/4\pi\phi$ of the photon beam. 
Explicit expression for $B$ reads 
\begin{eqnarray}
B_{\omega}
=
{\Delta N_{\omega} \over 
10^{3}\, (\Delta\omega/\omega)\,(2\pi)^2\,E_xE_y}\,
{I \over e}\,,
\label{eq.02}
\end{eqnarray}
Here $I$ is the electric current (in amperes) of the beam of particles, $e$ is the elementary charge.
The quantities  
$E_{x,y}=\left(\sigma^2+\sigma_{x,y}^2\right)^{1/2}\left(\phi^2+\sigma_{\phi_{x,y}}^2\right)^{1/2}$
stand for the total emittance of the photon source in the transverse directions.
Commonly, brilliance is measured  in
$\left[\hbox{photons/s/mrad}^{2}\hbox{/mm}^{2}/0.1\,\%\,\hbox{BW}\right]$.
To achieve this in~(\ref{eq.02}) one substitutes $I$ in amperes,
$\sigma,\,\sigma_{x,y}$ in millimeters, and  
$\phi,\, \sigma_{\phi_{x,y}}$ in milliradians.
If one uses the peak value of the current, $I_{\rm peak}$, then the corresponding
quantity is called \textit{peak brilliance}, $B_{\rm peak}$.

The number of photons within the BW 
is proportional to the spectral distribution $\d E (\theta\leq \theta_0)/\d(\hbar\om)$
of the energy radiated per particle in the solid angle $\Delta\Om \approx 2\pi \theta_0^2/2$ 
(the emission cone $\theta_0$ is assumed to be small, $\theta_0 \ll 1$): 
$\Delta N_{\om} = \left(\d E (\theta\leq \theta_0)/\d(\hbar\om)\right) \Delta\om/\om$.
Using this relation
one writes the brilliance in terms of the spectral distribution:
\begin{eqnarray}
B_{\om}
=
{\d E (\theta\leq \theta_0)\over \d(\hbar\om)}
\,
{1.58\times 10^{14}\, I\over E_x E_y}\,.
\label{eq.03}
\end{eqnarray}

\begin{figure*}[ht!]
\centering      
\includegraphics[scale=0.31,clip]{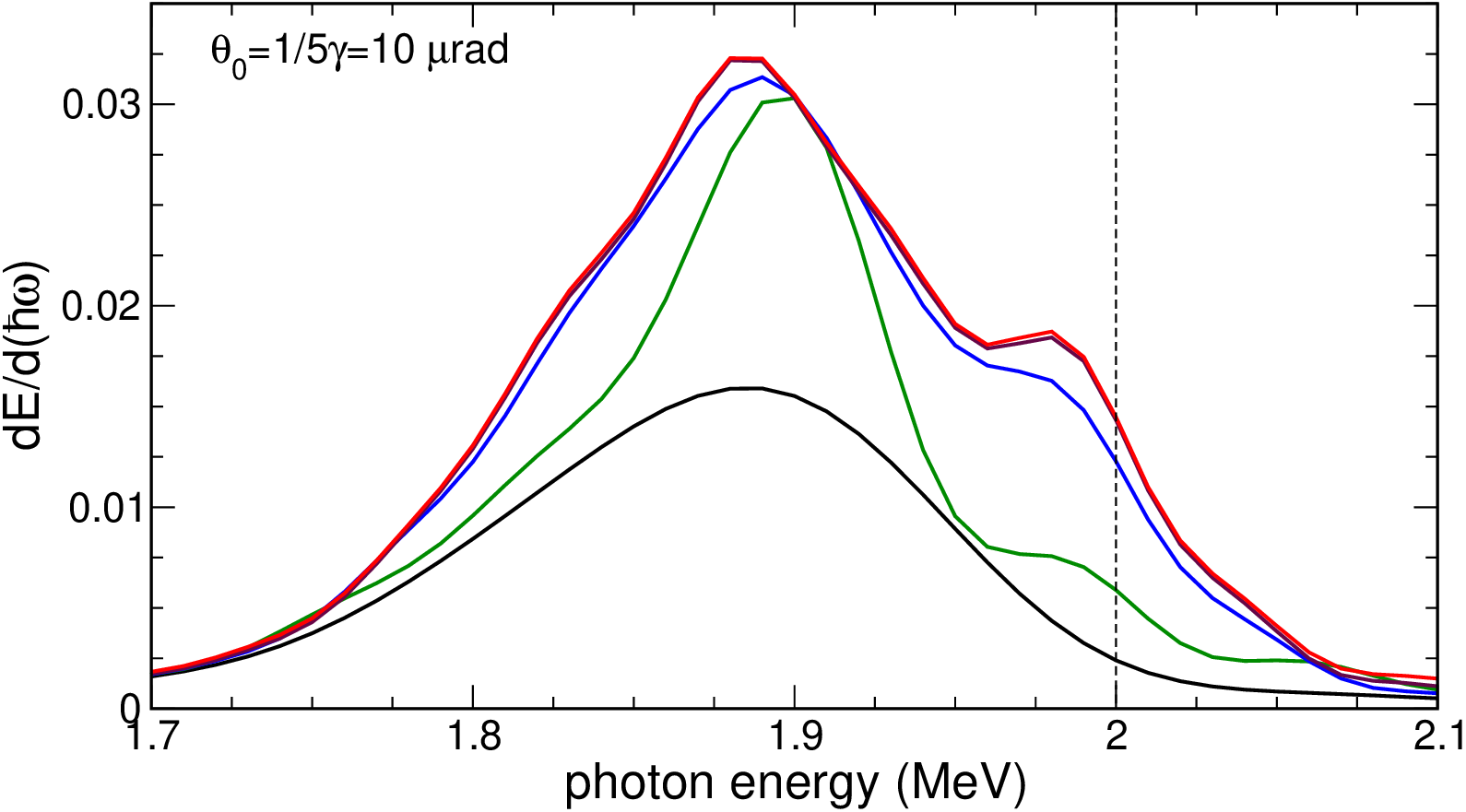}
\hspace*{0.5cm}
\includegraphics[scale=0.31,clip]{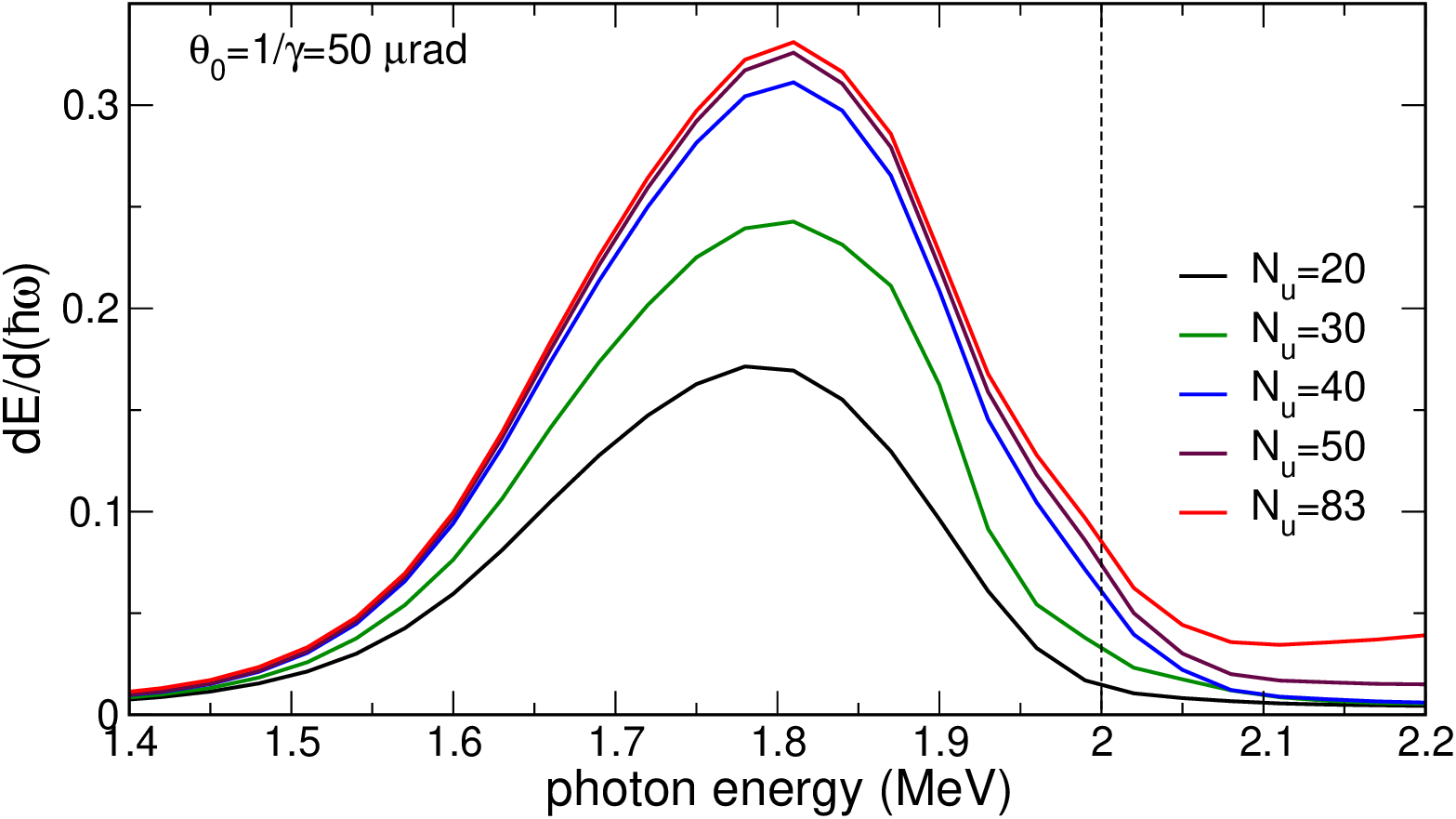}
\caption{
Spectral distribution $\d E (\theta\leq \theta_0)/ \d(\hbar\om)$ 
emitted within the cones $\theta_0=(5\gamma)^{-1} = 10$ $\mu$rad (left panel) and 
$\gamma^{-1} = 50$ $\mu$rad (right panel) corresponding to different number of undulator periods
$\Nu$ as indicated in the common legend in the right panel.
In both panels, a dashed line marks the value $\hbar\om_0=2$ MeV which is the
the position of the on-axis first harmonic peak as follows from Eq. (\ref{eq.01})
}
\label{Figure01.fig}
\end{figure*}

In the simulations, the spectral distribution $\d E_j (\theta\leq \theta_0)/ \d(\hbar\om)$
has been calculated for each trajectory simulated
($j=1,\dots, N_0$ with the total number of trajectories $N_0\approx 3\times10^3$).
Initial conditions at the crystal entrance (the transverse coordinates and velocities) 
have been generated using the normal distributions with the deviations $\sigma_{x,y}$ and $\sigma_{\phi_{x,y}}$
quoted above.
The resulting spectral distribution used to calculate the brilliance (\ref{eq.03}) 
has been obtained by averaging the individual spectra:
$\d E(\theta\leq \theta_0)/ \d(\hbar\om)=N_0^{-1}\sum_{j}\d E_j (\theta\leq \theta_0)/ \d(\hbar\om)$.
The sum is carried out over all simulated trajectories, and thus, its takes into account the contribution 
of the channeling segments as well as of those corresponding to the non-channeling regime.
More details on the simulation procedure as well on the formalism behind it one finds in the review
paper \cite{KorolSolovyov:EPJD_Colloquium_2021}.

\textcolor{black}{
As a case study below the emission of radiation from the CU with
the parameters labeled as ’Set (I) in Table \ref{Table01} is considered in more detail.
}

Figure \ref{Figure01.fig} presents the spectral distribution of radiation 
emitted within the narrow cone $\theta_0=(5\gamma)^{-1} = 10$ $\mu$rad (left) and
the wider one  $\theta_0=\gamma^{-1} = 50$ $\mu$rad (right).
The calculations were performed for different number of undulator periods
as indicated in the common legend shown in the right graph.

Several features of the computed spectra are to be noted.
First, all peaks are red-shifted from the estimate $\hbar\om_0 =2$ MeV that follows from Eq. (\ref{eq.01}).
Second, for large number of periods the spectrum becomes virtually independent on
$\Nu\gtrsim 50$.

To provide an explanation to both of these features we first note that in the vicinity of maximum 
the emission spectrum is mainly formed by the particles, which move in the channeling mode through 
the whole crystal (see Figure S1 in SM).
Therefore, the evolution of the spectrum is related, to a great extent, to the dynamics of 
\textit{channeling} particles in the course of their propagation through the crystalline medium.
Eq. (\ref{eq.01}) describes accurately the on-axis frequency of the fundamental harmonic in an 
ideal planar undulator, in which a particle moves along a perfect cosine trajectory 
$y(z)=a\cos(2\pi z/\lamu)$.  
In a CU, a channeling particle experiences (i) channeling oscillations while moving along a periodically
bent centerline, and (ii) stochastic motion along the $x$ axis due to the multiple scattering from  
crystal constituents.
The channeling oscillations lead to the following modification of the undulator parameter: $ K^2\to K^2 + K_{\rm ch}^2$.
Here $K_{\rm ch}\propto 2\pi \gamma  a_{\rm ch}/\lambda_{\rm ch}$ with $a_{\rm ch} \leq d/2$ and 
$\lambda_{\rm ch}$ being the amplitude and period of the channeling oscillations.
In the case of positron channeling, assuming harmonicity of the oscillations, one carries out 
averaging over the allowed values of $a_{\rm ch}$ and finds  
$\langle K_{\rm ch}^2 \rangle = 2 \gamma U_0 /3mc^2$ \cite{ChannelingBook2014}.
For $\E=10$ GeV in diamond(110) ($U_0\approx 20$ eV) this results in 
$\langle K_{\rm ch}^2 \rangle \approx 0.56$.
The stochastic motion along the $x$ axis results in a gradual increase in the rms 
scattering angle $\langle \theta_x^2 \rangle$ with the penetration distance $z$.
Hence, the motion of the particle is not restricted to the $(yz)$ plane and its emission within the cone centered along 
the incident beam direction becomes off-axis.
As a result, the denominator of the fraction in (\ref{eq.01}) increases,
$1 + K^2/2 \to 1 + K^2/2  + \langle K_{\rm ch}^2\rangle/2  + \gamma^2\langle \theta_x^2\rangle(z)  + \gamma^2\sigma_{\phi_x}^2$,
leading to the decrease in the fundamental harmonic frequency.
The increase in the rms scattering angle leads to the saturation of the emission spectrum with 
the crystal thickness.
Indeed, taking into account that an ultra-relativistic particle radiates, predominantly, within the cone $1/\gamma$ along
its instant velocity, the emission within the cone $\theta_0$ along the $z$ axis becomes negligibly small 
at the penetration distances $\tilde{z}$ when the relation 
$\left(\langle \theta_x^2\rangle(\tilde{z})  + \sigma_{\phi_x}^2\right)^{1/2} - \gamma^{-1} \gtrsim \theta_0$  
becomes valid. 
Here the term $\sigma_{\phi_x}^2$ accounts for the beam divergence at the crystal entrance.
The spectrum saturates at $L\sim \tilde{z}$.

The increase in the multiple scattering angle is also the main reason for a suppression 
of the simulated intensity as compared to the
ideal undulator, see Fig. S2 in SM.

\begin{figure*}[ht]
\centering      
\includegraphics[scale=0.32,clip]{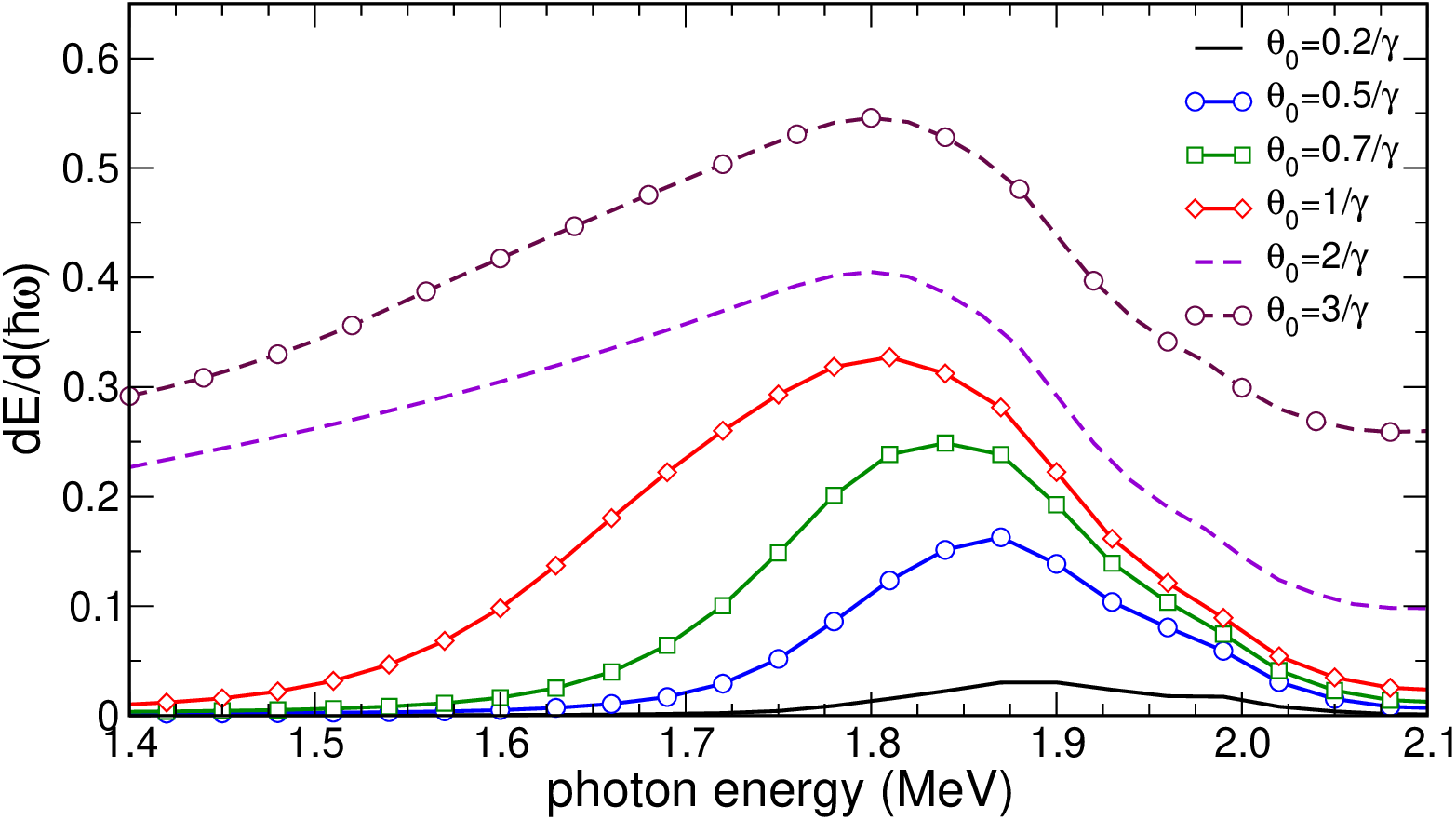}
\hspace*{0.5cm}
\includegraphics[scale=0.32,clip]{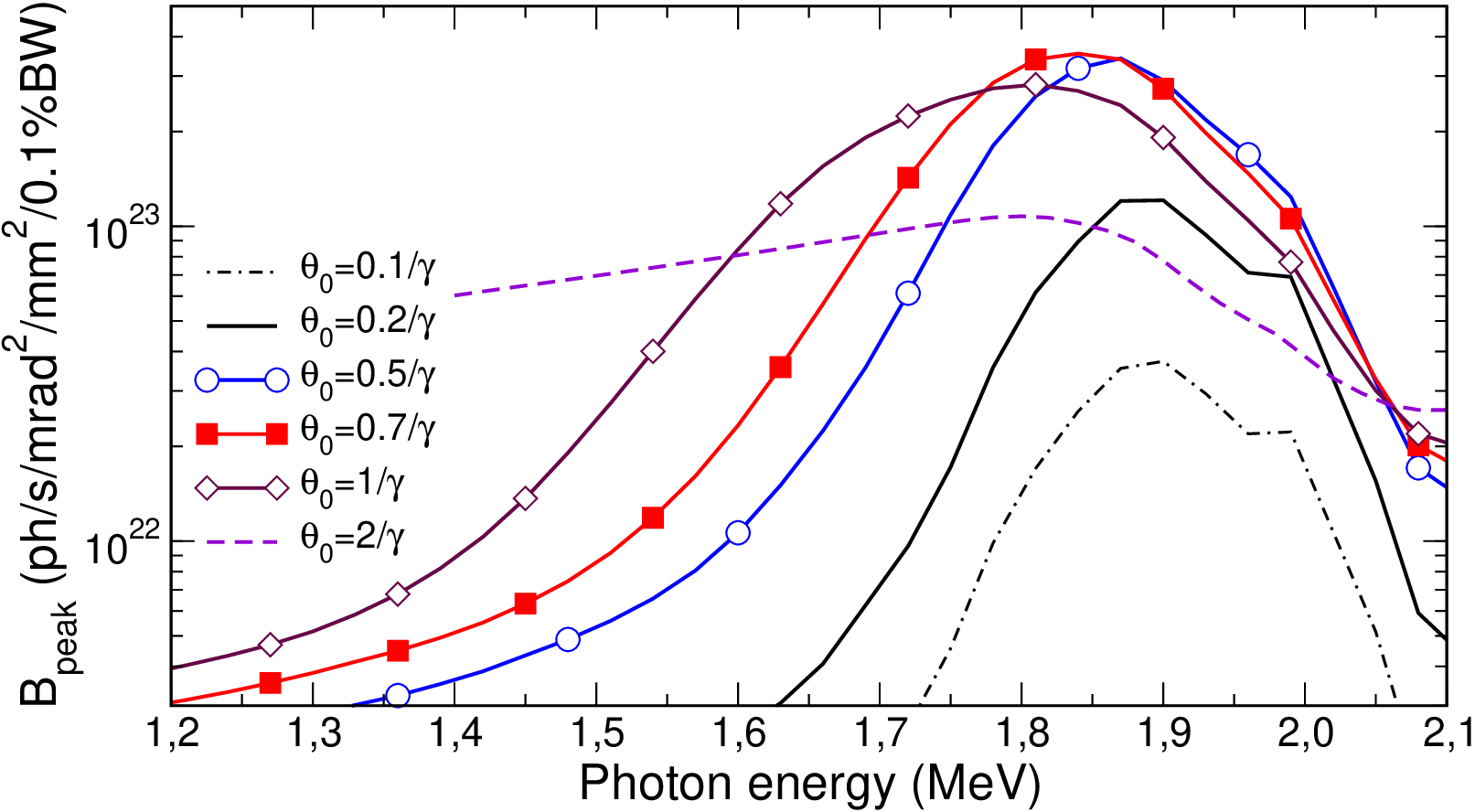}
\caption{
\textcolor{black}{
Spectral distribution
of radiation emitted (left) and
peak brilliance of CU-LS (right)
calculated for different emission cones $\theta_0$ as indicated.
Both panels refer to the CU with
the parameters labeled as ’Set (I) in Table \ref{Table01}.
}
}
\label{Figure02.fig}
\end{figure*}

\textcolor{black}{
The CU radiation spectra calculated for different emission cones, $\gamma\theta_0=0.1-4$, are presented in
the left graph of Fig. \ref{Figure02.fig}.
All curves correspond to the crystal length $L=5.1$ mm ($\Nu=60$).
These data supplemented with the aforementioned values for the beam size, divergence and 
peak current allow one to calculate the peak brilliance of the CU-LS, Eq. (\ref{eq.03}).
The results of calculations are shown in Fig. \ref{Figure02.fig} \textit{right}.
It is seen that in contrast to $\d E(\theta\leq \theta_0)/ \d(\hbar\om)$, which is an
increasing function of the emission cone, the peak brilliance is a non-monotonous function
due to the presence of the terms proportional to $\theta_0$ in the total emittance $E_{x,y}$ that 
enter the denominators in Eqs. (\ref{eq.02}) and (\ref{eq.03}).
In the case study considered here the maximum value }%
$B_{\rm peak}^{\max}\approx 3.5\times10^{23}$ {photons/s/mrad$^2$mm$^2$/0.1\,\%\,{BW}
 \textcolor{black}{
is achieved at
$\hbar\om=1.85$ MeV for $\theta_0=0.7/\gamma = 35$ $\mu$rad.}

\begin{figure}[ht]
\centering      
\includegraphics[scale=0.32,clip]{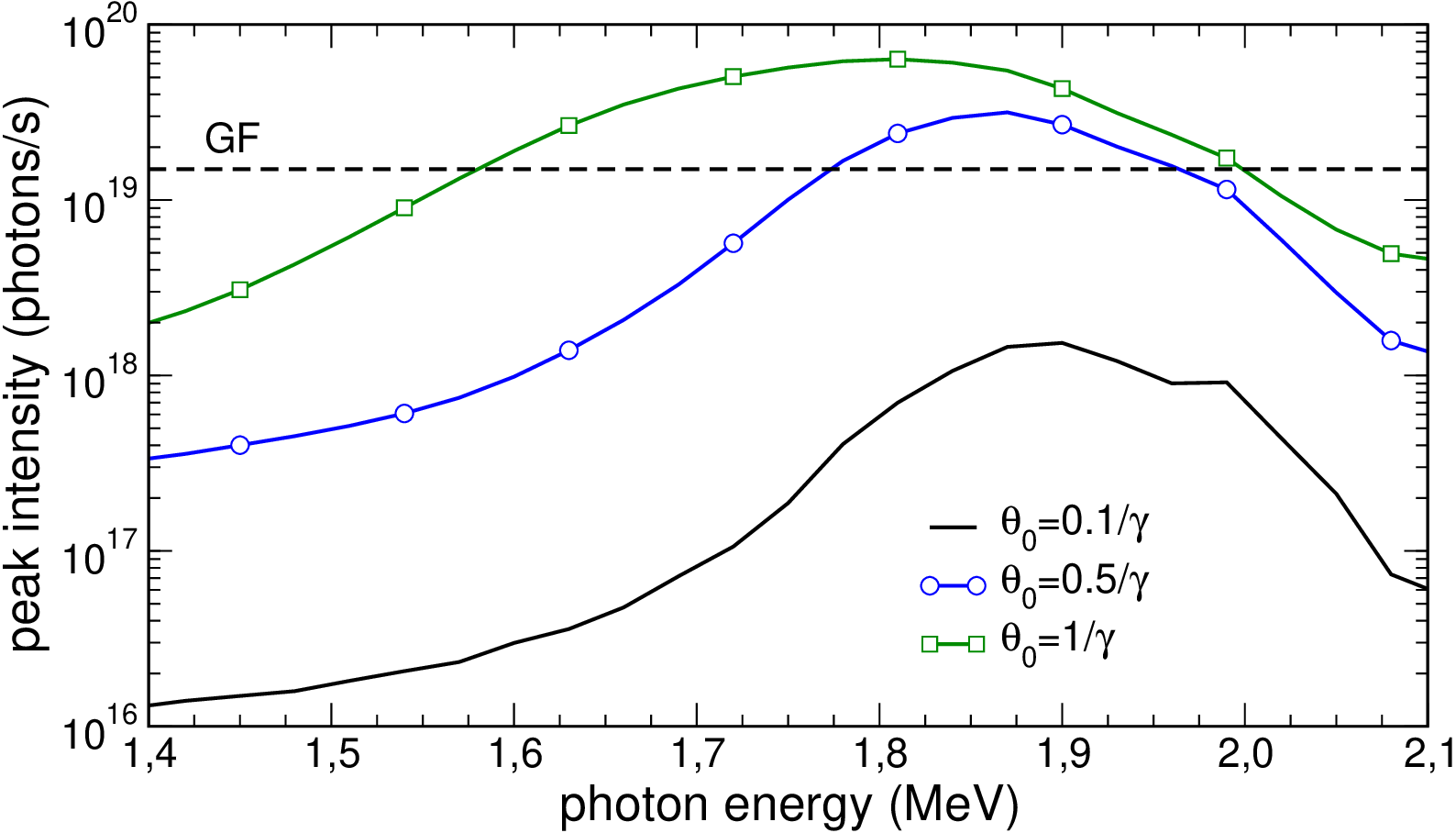}%
\caption{
\textcolor{black}{
Peak intensity (number of photons per second)
of the CU-LS with the parameters indicated in
’Set (I) in Table \ref{Table01}.
The horizontal dashed line marks the peak intensity
of the LS discussed in the Gamma Factory (GF) proposal for CERN \cite{Krasny:2018xxv}.
The intensities of the laser-Compton backscattering LSs, being on the
level $10^{13}-10^{14}$ photon/s
\cite{Rehman_EtAl_ANE_v105_p150_2017}, are not shown.
}
}
\label{Figure03.fig}%
\end{figure}

\textcolor{black}{
The product $\Delta N_{\omega} I /e$ on the right-hand side of Eq. (\ref{eq.01})
defines the number of photons per unit time interval (intensity)
emitted in the solid angle $\Delta \Om$ and frequency
interval $\Delta \om$.
Using the peak current $I_{\rm peak}=3.1$ kA 
one calculates the peak intensity achievable in the CU with given bending parameters 
exposed to the FACET-II beam. 
Figure \ref{Figure03.fig} shows the corresponding dependences calculated for
for the BW $\Delta \om /\om = 0.01$ and for several  emission cones as indicated. 
}

\textcolor{black}{
It is instructive to compare the CU-LS intensity
 with the intensity predicted in the Gamma Factory (GF) proposal for CERN
\cite{Krasny:2018xxv,GammaFactory:Letter_of_intent2019}.
The latter discusses a concept of the LS based on the resonant
absorption of laser photons by ultra-relativistic ions.
It is expected that the intensity of the GF LS will be orders of magnitude
higher that the presently operating laser-Compton LSs aiming at the
value of $\langle I_{\rm GF} \rangle =10^{17}$ photons/s
in the gamma-ray domain above 1 MeV.
The quoted value of the intensity refers to the average beam current.
To calculate the corresponding peak value one multiplies
$\langle I_{\rm GF}\rangle$ by a factor
$\approx 100/0.64 \approx 150$, which is the ratio of the bunch spacing ($\approx$100 ns) to the bunch length  ($\approx$0.64 ns).
The dashed line in Fig. \ref{Figure03.fig} indicates the peak intensity of the GF LS
equal to $I_{\rm GF}=1.5\times10^{17}$ photons/s.
To be noted is that for the bending amplitude and period considered
the intensity of the CU-LS in the vicinity of its first harmonic emitted within
the cones greater than $\approx 0.4/\gamma=20$ $\mu$rad exceeds the
the peak value $I_{\rm GF}$.
}

\begin{figure} [h]
\centering
\includegraphics[scale=0.32,clip]{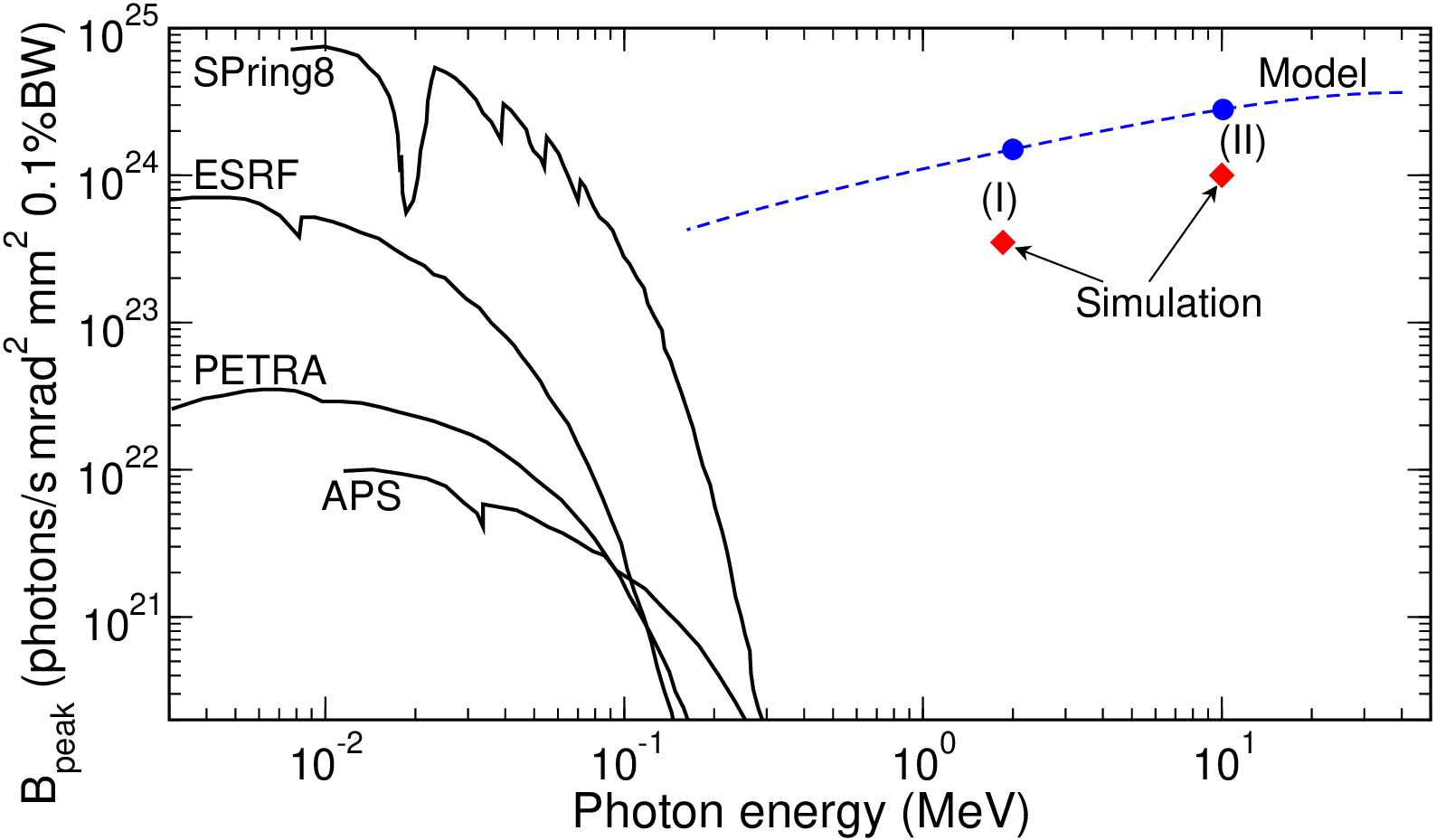}
\caption{
\textcolor{black}{
Comparison of the peak brilliance available at several synchrotron radiation facilities (APS, ESRF, PETRA,
SPring8) with that for the CU-LS.
The filled diamond symbols show the brilliances that correspond to Sets (I) and (II)
 in Table \ref{Table01}.
The dashed line stands for the model estimation \cite{KorolSolovyov:EPJD_CLS_2020};
the filled circles marked the estimated brilliances for Sets (I) and (II).
}
}
\label{Figure_04.fig}%
\end{figure}

\textcolor{black}{
Figure \ref{Figure_04.fig} compares peak brilliance $B_{\rm peak}(\om)$ of several operational synchrotron
radiation facilities with that achievable by means of the diamond-based CU exposed to the FACET-II positron beam.
The dashed line shows the results of the model calculations \cite{KorolSolovyov:EPJD_CLS_2020} that have provided,
for each photon energy $\hbar\om$, the highest value of $B_{\rm peak}(\om)$ by scanning through the ranges of $a$ and $\lamu$.
Symbols show the values of $B_{\rm peak}(\om)$ obtained for Sets (I) and (II) of the parameters indicated in Table \ref{Table01}:
the results of model calculations estimations (circles) are to be compared with the
results of the accurate numerical simulations (diamonds).
}

\textcolor{black}{
The results of accurate numerical simulations presented in this Letter refer to the specific case study
of the radiation emission by a 10 GeV positron beam (with the transverse size and divergence indicated in 
Ref. \cite{FACETII_Conceptual_Design_Rep-2015}) channeling in an oriented periodically bent diamond (110) crystal.
The parameters indicated in Table \ref{Table01} maximize the peak brilliance of radiation in the vicinity of the $\hbar\om=2$ MeV (Set I) and 10 MeV (Set II) as it has been
estimated by means of the model approach \cite{KorolSolovyov:EPJD_CLS_2020}.
The simulations demonstrated that the peak brilliance of the CU-LS is comparable to or even higher than that
achievable in conventional synchrotrons in the much lower photon energy range.
The intensity of radiation exceeds the values predicted in the GF proposal for CERN.
}

By tuning the bending amplitude and period one can maximize brilliance for given parameters of
a positron beam and/or chosen type of a crystalline medium. 
As a result extremely high values of brilliance can be achieved in the photon energy range 
$10^1 \dots 10^2$ MeV by currently available (or planned to be available
in near future) positron beams \cite{KorolSolovyov:EPJD_CLS_2020}.
For each set of the input parameters (beam energy and emittance, bending amplitude and period, 
crystal type and thickness, detector aperture etc.)
to provide accurate data on the brilliance from a CLS rigorous numerical simulations, similar to those 
presented in this Letter, must be carried out. 
When doing this, the results of the model approach can be used as the initial estimates of the  
the ranges of the parameters to be used in the simulations.

It is worth mentioning that the size and the cost of CLSs are orders of
magnitude less than those of
modern LSs based on the permanent magnets. 
This opens many practical possibilities for the efficient generation of gamma-rays with various intensities and 
in various ranges of wavelength by means of the CLSs on the existing and newly constructed
beam-lines.

\vspace*{0.2cm}
The work was supported in part by the DFG Grant (Project No. 413220201) and by the H2020 RISE-NLIGHT project (GA 872196).
We acknowledge the Frankfurt Center for Scientific Computing (CSC) for providing computer facilities.


\end{document}